\documentclass{emulateapj}
\usepackage{color}

\newcommand{\msun}{$\mathrm{M_{\odot}}$}
\newcommand{\kms}{$\mathrm{km\,s^{-1}}$}

\newcommand{\drvm}{$\Delta \mathrm{RV_{max}}$}
\newcommand{\rvet}{$\delta \mathrm{RV_{thr}}$}
\newcommand{\mym}{$\mathrm{mergers\,yr^{-1}\,M_{\odot}^{-1}}$}
\newcommand{\ym}{$\mathrm{yr^{-1}\,M_{\odot}^{-1}}$}
\newcommand{\fb}{$f_{\rm bin}$}

\newcommand{\sw}{SWARMS}

\slugcomment{Draft Version \today}

\shorttitle{White Dwarf Merger Rate}
\shortauthors{Badenes \& Maoz}
\begin{document}

\title{The Merger Rate of Binary White Dwarfs in the Galactic Disk}

\author{Carles Badenes\altaffilmark{1,2,3} and Dan Maoz \altaffilmark{2}}

\altaffiltext{1}{Department of Physics and Astronomy and Pittsburgh Particle Physics, Astrophysics, and Cosmology Center
  (PITT-PACC), University of Pittsburgh, 3941 O'Hara Street, Pittsburgh, PA 15260, USA; badenes@pitt.edu}

\altaffiltext{2}{School of Physics and Astronomy, Tel-Aviv University, Tel-Aviv 69978, Israel; maoz@astro.tau.ac.il}

\altaffiltext{3}{Benoziyo Center for Astrophysics, Weizmann Institute of Science, Rehovot 76100, Israel}

\begin{abstract}
  We use multi-epoch spectroscopy of $\sim$4000 white dwarfs in the Sloan Digital Sky Survey to constrain the properties of the
  Galactic population of binary white dwarf systems and calculate their merger rate. With a Monte Carlo code, we model the
  distribution of \drvm, the maximum radial velocity shift between exposures of the same star, as a function of the binary
  fraction within 0.05 AU, \fb\, and the power-law index in the separation distribution at the end of the common envelope phase,
  $\alpha$.  Although there is some degeneracy between \fb\ and $\alpha$, the the fifteen high \drvm\ systems that we find
  constrain the combination of these parameters, which determines a white dwarf merger rate per unit stellar mass of
  $1.4^{+3.4}_{-1.0} \times10^{-13}$ \ym\ (1$\sigma$ limits). This is remarkably similar to the measured rate of Type Ia
  supernovae per unit stellar mass in Milky Way-like Sbc galaxies. The rate of super-Chandrasekhar mergers is only
  $1.0^{+1.6}_{-0.6}\times10^{-14}$ \ym. We conclude that there are not enough close binary white dwarf systems to reproduce the
  observed Type Ia SN rate in the `classic' double degenerate super-Chandrasekhar scenario. On the other hand, if
  sub-Chandrasekhar mergers can lead to Type Ia SNe, as has been recently suggested by some studies, they could make a major
  contribution to the overall Type Ia SN rate. Although unlikely, we cannot rule out contamination of our sample by M-dwarf
  binaries or non-Gaussian errors. These issues will be clarified in the near future by completing the follow-up of all 15 high \drvm\
  systems.

\end{abstract}


\section{INTRODUCTION}
\label{sec:Intro}

The nature of the progenitor systems of Type Ia supernovae (SN Ia) remains one of the key open issues in stellar evolution. There
is a general agreement that the exploding star is a CO white dwarf (WD) that is somehow ignited following accretion of material
from a binary companion, but the identity of this companion is still a matter of debate. Most theoretical scenarios for SN Ia
progenitors can be divided into two broad classes: single degenerate (SD) systems \citep{whelan73:SNI_SD}, where the companion is
a non-degenerate star, and material is transferred over $\lesssim$ 1 Myr via Roche lobe overflow or wind accretion, and double
degenerate (DD) systems \citep{iben84:typeIsn,webbink84:DDWD_Ia_progenitors}, where the companion is another WD, and mass transfer
happens over much shorter time scales in a merger event.

Recent developments have provided some evidence in favor of the DD scenario. Sensitive searches have failed to find signs of a
mass-losing nondegenerate companion in radio observations \citep{HoreshAssaf2011,Chomiuk2012}, early light curves
\citep{Hayden2010,Bianco2011,BloomJoshuaS.2011} and supernova remnants \citetext{\citealp{badenes07:outflows,Schaefer2012}, but
  cf. \citealp{Williams2011}, and see \citealp{SternbergA.2011} for some statistical evidence of absorption in SN Ia spectra that
  could be of SD origin}. Several independent measurements of the delay time distribution (the specific rate of SN Ia as a
function of time after a hypothetical brief burst of star formation) have converged onto a $\sim t^{-1}$ shape extending from a
few hundred Myr to several Gyr \citep[see][for a recent review]{MaozDan2011}, a form expected from a population of DD systems that
merge due to gravitational wave emission.

In this Letter, we present the first measurement of the local WD merger rate and use it to test the viability of the DD scenario
for SN Ia progenitors. To date, orbital parameters have been measured for over 40 individual DD systems with periods ranging
between 12 minutes \citep{Brown2011b} and several days \citep[e.g.][]{MarshT.R.1995,nelemans05:SPY_IV,kilic10:WD_binaries_merge},
but the fundamental properties of the Galactic DD population are still poorly known. Some small WD samples have been used to
estimate parameters like the binary fraction \citep[][find it to be between $0.017$ and $0.19$, based on 46 WDs]{Maxted1999a} or
the merger rate limited to systems that contain extremely low mass ($\leq 0.25$ \msun) WDs \citep[$\sim 4 \times 10^{-5} \,
\mathrm{yr^{-1}}$ based on 12 systems,][]{Brown2011}. Here, we adopt a \textit{statistical} approach to characterize the DD
population, using the unique capabilities of the Sloan Digital Sky Survey (SDSS) for time resolved spectroscopy of thousands of
WDs. We use this large data set to measure the distribution of \drvm, the maximum radial velocity (RV) shift between exposures of
the same WD. In a companion paper (Maoz, Badenes \& Bickerton 2012, henceforth Paper I), we describe how this distribution
constrains the binary fraction and the separation distribution. Because the evolution of detached DD systems is driven only by
gravitational wave emission
%\citep{Peters1963}, 
these parameters, together with the WD masses, uniquely determine the local merger rate,
which can be compared to measurements of the specific SN Ia rate in nearby Milky Way-like galaxies.

\section{OBSERVATIONS}

The SDSS \citep{york00:SDSS_Technical} contains the largest available collection of WD spectra
\citep{Kleinman2004,eisenstein06:WD_SDSS_DR4}. The latest version of the SDSS WD catalog, corresponding to DR7, has over $17000$
entries \citep{Kleinman2009}\footnote{The pre-publication catalog, July 2010 version, was kindly provided by S. Kleinman (private
  communication), and has $17,371$ entries.}.  For simplicity, we have restricted our analysis to non-magnetic DA WDs that show no
obvious signs of main sequence companions, and have no confirmed or possible absorption lines from any elements other than H - i.e.,
objects classified as `DA' or `DA:' in the catalog,
%and no features characteristic of cool main sequence stars in the red part of their
%spectra. 
%We cross-matched this sample with the catalog of WDs with main sequence companions recently published by
%\cite{rebassa10:SDSS_WD+M_catalog}, and removed 7 binaries listed there that were missed in the DR7 catalog, which left a
%total of $12763$ WDs.
We further removed from this sample seven WDs with main sequence companions listed by \cite{rebassa10:SDSS_WD+M_catalog}, leaving
$12763$ objects.

\begin{figure*}

  \centering
 
  \includegraphics[angle=90,scale=0.63]{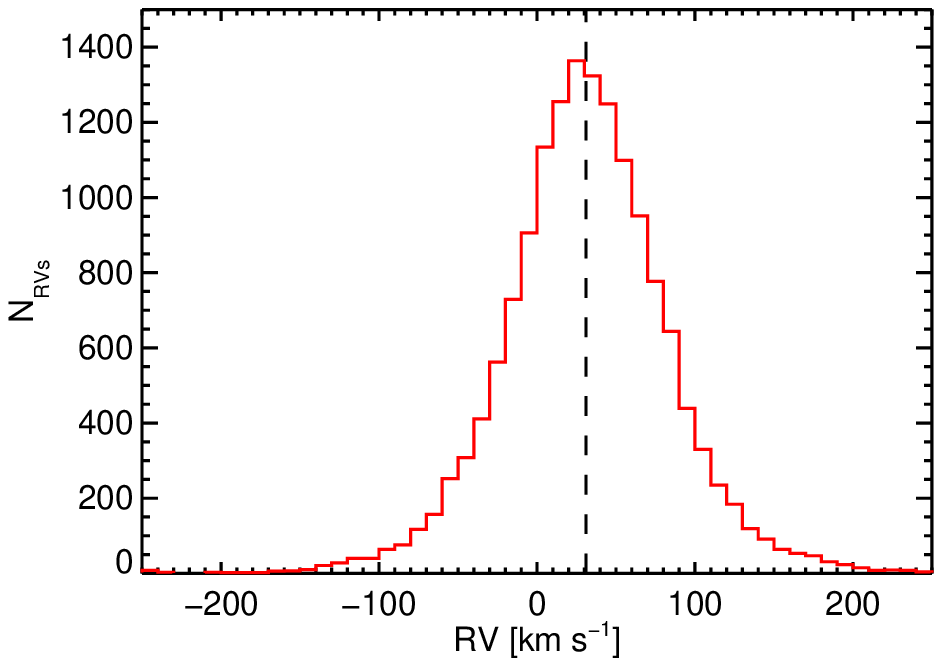}
  \includegraphics[angle=90,scale=0.63]{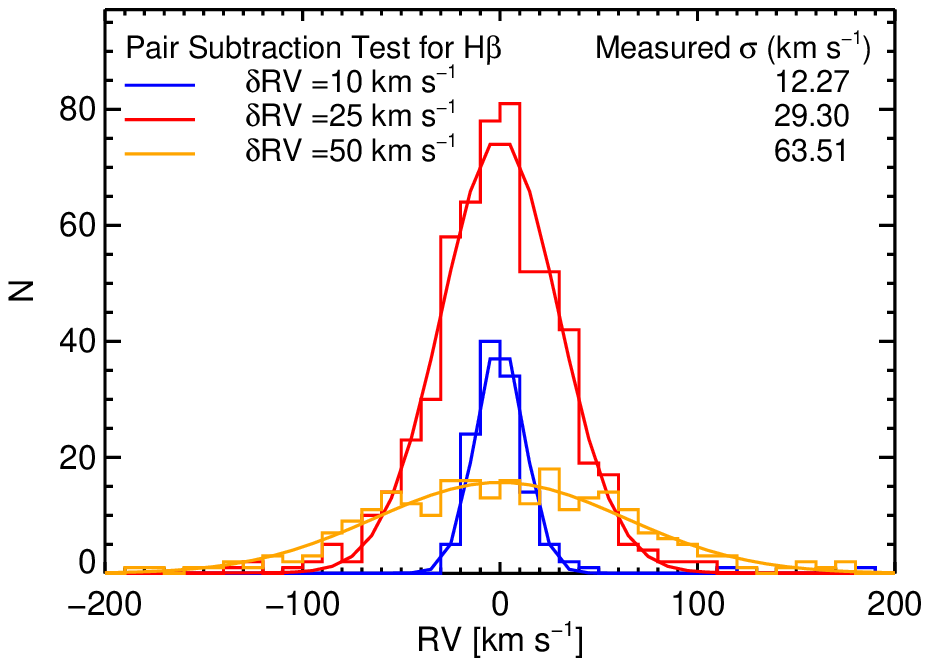}
  \includegraphics[angle=90,scale=0.63]{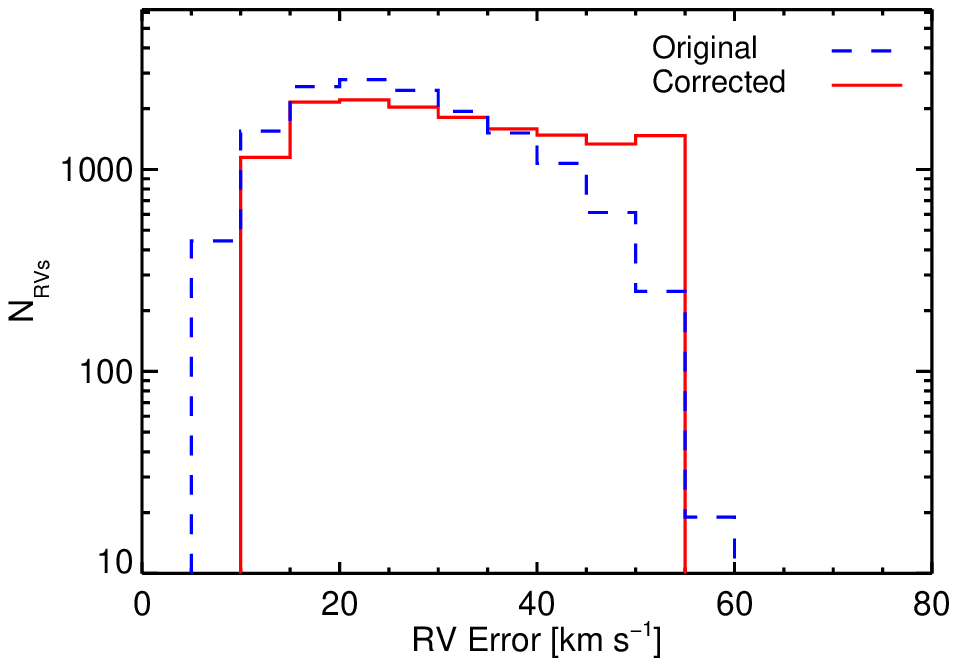}

  \caption{\textit{Left:} Distribution of RVs. The nonzero mean ($31.0\pm0.4$ \kms, dashed line) is due
    to gravitational redshift.  \textit{Center:} Illustration of the pair subtraction test used to correct the errors
    reported by MPFIT (in this example, for H$\beta$). \textit{Right:} Distribution of $\delta$RV before (dashed blue
    line) and after (solid red line) correction. } \label{fig-RVDist}

\end{figure*}

Like all SDSS spectra, the spectra of these WDs were divided into three or more sub-exposures to facilitate cosmic ray rejection
\citep{stoughton02:SDSS_EDR}. Since DR7 \citep{abazajian09:SDSS_DR7}, these sub-exposures are available from the SDSS server. The
ongoing \sw\ survey is using these sub-exposures to identify short-period binary WD candidates in SDSS, which are then followed up
\citep{badenes09:SWARMS_I,mullally09:DDWDs}. The DS/DT collaboration \citep{BickertonSteven2011} has implemented a pipeline to
handle cosmic ray rejection and derive stable wavelength solutions for all sub-exposures of the same object. This Letter presents
the first results obtained using the DS/DT pipeline -- see \cite{BickertonSteven2011}
%and forthcoming DS/DT publications 
for technical details.

To measure RVs, we normalize each spectrum, dividing it by a highly smoothed version of itself, and we fit four Balmer lines --
H$\alpha$ through H$\delta$ -- with Voigt profiles in absorption.
% \citep[a Lorentzian line with a Gaussian
% core,][]{thompson04:PY_Vul}. 
We find the best-fit model for each line using MPFIT \citep{markwardt09:MPFIT}, an IDL implementation of the Levenberg-Marquardt
algorithm. For each sub-exposure $n$, these fits yield four RV measurements RV$_{n,i}$, and four 1$\sigma$ errors
$\delta$RV$_{n,i}$ ($i=0..3$). To define a single RV for each sub-exposure, we impose a threshold on the error of individual RV
measurements, \rvet, and discard values with larger errors or that deviate by $>1 \sigma$ from the weighted mean, RV$_{n}= \left(
  \sum_{i} w_{n,i} \times \mathrm{RV}_{n,i} \right) / \sum_{i} w_{n,i}$, with $w_{n,i} = 1/\delta \mathrm{RV}_{n,i}$. In
sub-exposures with at least \textit{two} surviving RVs, we re-calculate the weighted mean RV$_{n}$, and define its error as
$\delta$RV$_{n}=\sqrt{ 1/ \sum_{i} w_{n,i}^2}$. Finally, we define \drvm\ as the difference between the highest and lowest
RV$_{n}$ in each object.
%, \drvm$=\max(\mathrm{RV}_{n}) -
%\min(\mathrm{RV}_{n})$.
This procedure eliminates noisy sub-exposures with internally inconsistent RVs, effectively culling most WDs fainter than $g\sim$
19.  We find that \rvet=80 \kms\ is a good compromise between sample size and quality, with 4063 WDs and 15,236 RV
measurements. The distribution of RVs and $\delta$RVs in this sample is shown in Figure~\ref{fig-RVDist}. The mean of the RV
distribution, $31.0 \pm 0.4$ \kms, is non-zero and positive due to gravitational redshift at the WD surfaces. This is the expected
value for a local WD population with an average mass of $\sim0.6$ \msun, and confirms the $32.6 \pm 1.2$ \kms\ redshift obtained
by \cite{Falcon2010a} using a different sample of $449$ DA WDs observed at high resolution \citetext{$\sim$16 \kms\ per pixel, see
  \citealp{napiwotzki01:SPY_survey} -- for comparison, SDSS spectra are $\sim$ 70 \kms\ per pixel}. The agreement shows that our
procedure yields well-calibrated RVs.

A key ingredient in the Monte Carlo models described in Section \ref{sec:drvm-distribution} and Paper I is the distribution of
errors, $\delta$RV. We therefore estimate the errors independently, and compare them to the errors reported by MPFIT. For each
Balmer line of each WD, we find the difference between random pairs of RVs that have similar reported values of $\delta$RV. These
differences have an approximately Gaussian distribution, with a $\sigma$ that gives an empirical estimate of the true error in the
RV measurements, times $\sqrt 2$. Using this procedure, we find that MPFIT tends to underestimate the RV errors. We obtain the
corrected $\delta$RV distribution shown in Figure \ref{fig-RVDist} by running a grid of such pair subtraction tests.

\section{THE WD MERGER RATE}
\label{sec:drvm-distribution}

\begin{figure}[!t]

  \centering
  
  \includegraphics[angle=90,scale=0.9]{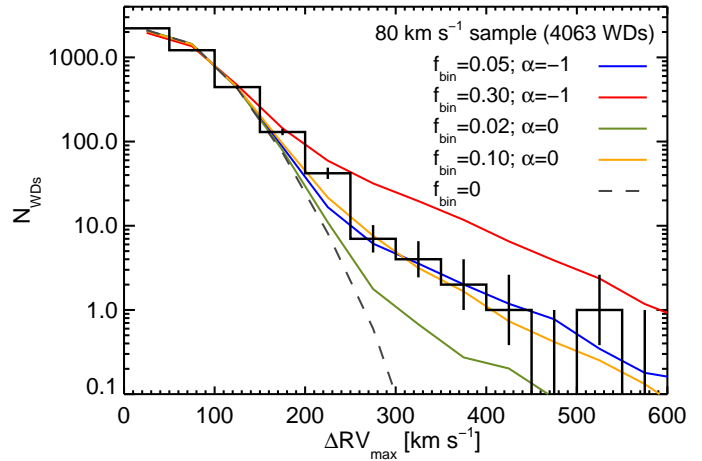}
  
  \caption{Distribution of \drvm\ with Poisson error bars (black histogram), compared
    to model distributions (blue, red, orange and green solid curves). The dashed black curve is a model with
    \fb=0.} \label{fig-DeltaRV_Models}

\end{figure}

The distribution of \drvm\ values in the \rvet=80 \kms\ sample is shown in Figure \ref{fig-DeltaRV_Models}. The vast
majority of WDs have low \drvm\ -- these are either isolated WDs or DD systems where the data show no evidence for
significant RV shifts, because the period is too long, or the inclination is too small, or the sub-exposures were taken
at similar orbital phases. However, there is a significant tail of WDs with high \drvm, extending beyond 500 \kms. From
the distribution of SDSS sampling times and RV errors, we do not expect any non-binary WDs to have \drvm\ higher than
$\sim$300 \kms\ (dashed line in Figure \ref{fig-DeltaRV_Models} -- see Paper I for details). The eight objects
with higher \drvm\ are therefore real binaries, detected at high significance. Some of these are already
published discoveries, like SDSS 1257 \citep[][\drvm$=538$ \kms]{badenes09:SWARMS_I}, or SDSS 0923 \citep[][\drvm$=410$
\kms]{Brown2010}. Others will be the subject of
forthcoming papers.

At intermediate values of \drvm, between $\sim$200 and 300 \kms\, follow-up observations are necessary to tell which of the
several dozen WDs are real binaries, but here we are only concerned with the statistics of the \drvm\ distribution, not the binary
character of individual systems. As a precaution, we have visually vetted the spectra of the 107 WDs with \drvm$\geq 175$ \kms\ to
check for any misclassified objects, WDs with weak line emission, spurious line fits, etc., and removed them from the
distribution.

The Monte Carlo techniques that we use to model the \drvm\ distribution are described in detail in Paper I. The method involves
generating a large number of WD systems with a given binary fraction, drawing the binary separation, $a$, and the component masses
from given distributions, assigning random inclinations and phases, choosing the photometric primary, and folding the resulting RV
curves through the observational parameters (sampling times and $\delta$RV distribution) of the SDSS data. Because we can only
detect binaries at high confidence with \drvm$\gtrsim250$ \kms, there is an upper limit to the separations that we can probe. For
an extreme-mass-ratio WD pair with $M_{1}=1.1$ \msun\ and $M_{2}=0.2$ \msun, an RV curve with a peak-to-peak amplitude of 250
\kms\ corresponds to a separation of $\sim$ 0.05 AU. Therefore, we define the measured binary fraction, \fb, as the fraction of
WDs with companions within $a_{\mathrm{max}}=0.05$ AU. As explained in Paper I, our definition of \fb\ does \textit{not} include
an additional population of extremely low-mass binaries with primaries below 0.25 \msun, which we assume are always in
short-period binaries \citep[see][and references therein]{Brown2011}. To constrain the distribution of binary separations, we
model it as a power-law with index $\alpha$ at the end of the final common-envelope episode, $n(a) \propto a^{\alpha}$. In Paper
I, we derive analytically the time-evolved form of this distribution due to gravitational wave emission and merger events,
integrated over the star-formation history of the Galaxy. This evolved separation distribution is used for generating the
simulated WD pairs. Primary masses are chosen from the observed distribution of isolated WDs in
\cite{kepler07:WD_mass_distribution}. Secondary masses are chosen from a power-law mass-ratio distribution of index $\beta$. In
practice, the results have a weak dependence on $\beta$ (see Paper I for details), and the analysis we present here assumes
$\beta=0$. We choose the photometric primary by random draw, except for $<0.35$ \msun\ secondaries, which we assume are always
photometric primaries (see Paper I).

% Independent constraints on \fb\ and $\alpha$ parameters are hard to find in the literature. The main sequence
% progenitors of the WDs we study here have have binary fractions between 40\% and 70\%, but the bulk of companions are
% found at wide separations \citep[most have $P>$1000 days, see][]{Raghavan2010}. For our definition of \fb\, \redpen{[say
%   something intelligent here]}. Little is known, theoretically or observationally, about the separation distribution at
% the time the WDs emerge from their final common envelope phase. A power-law distribution with a negative index close to
% $-1$ is often assumed, by analogy to what is observed for main sequence stars \citep[e.g,][]{Fischer1992,Poveda2007},
% but other forms are certainly possible. Some theoretical work seems to support the $\alpha \sim -1$ hypothesis
% \citep{Deloye2010}, although this remains a controversial subject \citep[see][for a recent review]{Ivanova2011}. Because
% the evolution of detached DD systems is driven exclusively by gravitational wave emission \citep{Peters1963}, the
% combination of $\alpha$ and \fb\ uniquely determines the merger rate per WD in the sample. This rate can then be
% translated to a specific merger rate using the known local space density of WDs \redpen{[I need the reference for
%   this]}.

Each combination of \fb\ and $\alpha$ leads to a predicted \drvm\ distribution in the survey.  Examples of such model
distributions are plotted alongside the observed one in Fig.~\ref{fig-DeltaRV_Models}. This plot showcases the discriminating
power of the \drvm\ distribution, in that some combinations of \fb\ and $\alpha$ are allowed by the data, and some are clearly ruled
out. The validity of each model can be quantified by multiplying the Poisson probabilities of finding the observed number of
systems in each \drvm\ bin, given the expectation from the model. In all comparisons between models and observations, we use only
the \drvm\ bins above 250 \kms\ (15 systems in total).

\begin{figure}[!t]

  \centering
  
  \includegraphics[angle=90,scale=0.9]{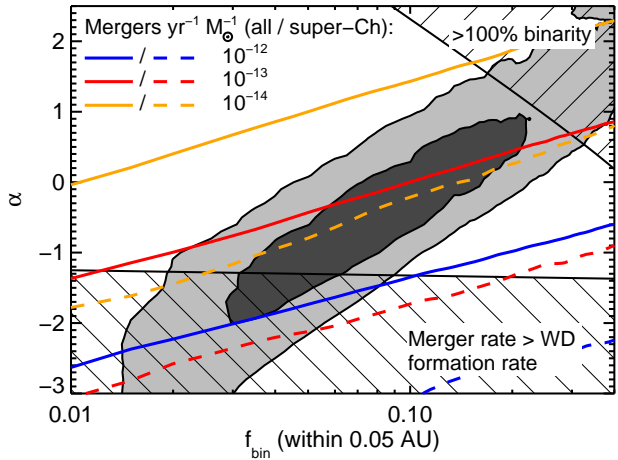}
  
  \caption{Likelihood contours in the \fb, $\alpha$ plane, indicating the $95\%$ (light gray) and $68\%$ (dark gray)
    confidence levels. The overlaid color lines are curves of constant WD merger rate, total (solid) and
    super-Chandrasekhar (dashed). The hatched regions are ruled out by requiring that less than 100\% of WDs have
    companions with P$<$12 days (upper right corner) and that WDs do not merge faster than they are formed (bottom part
    of the plane) -- see text for details.} \label{fig-LikCont}
  
\end{figure}

The likelihoods of different WD population models are shown in Figure \ref{fig-LikCont}, with the $95\%$ and $68\%$ confidence
level contours marked. Lower binary fractions require more negative values of $\alpha$ to reproduce the observations, and vice
versa. This is because the tail of WDs at high \drvm\ (i.e.  short-period binaries) can result either from WD populations where
many binaries are formed at all separations (high \fb, more positive $\alpha$), or few binaries are formed but with preferentially
small separations (low \fb, more negative $\alpha$). Some population models can be ruled out using simple arguments. We can
require, for instance, that $<$100\% of WDs have a companion with P$<$12 days (the circularization limit), which would be in
conflict with main sequence binary surveys \citep[e.g.][]{Raghavan2010}. We can also require that WDs do not merge faster than
they are formed, because we \textit{do} observe pre-merger systems. These two conditions rule out the striped regions in the \fb,
$\alpha$ plane marked in Figure \ref{fig-LikCont}.

Every simulated system has aknown merger time, and therefore it is straightforward to obtain the merger rate for each simulated
binary population, whether in total, or for specific cases, such as super-Chandrasekhar systems.  The merger rate per WD in the
sample can be converted to a merger rate per unit stellar mass in the solar neighborhood using measured estimates of the local WD
number density \citep{Sion2009} and stellar mass density \citep{McMillan2011}.  Curves of constant WD merger rate per unit stellar
mass are shown in Figure \ref{fig-LikCont}. As explained in Paper I, the curves appear as straight lines in $\alpha-\log(f_{\rm
  bin})$ space. The slope of the solid lines (all mergers) is $1/\log(a_{\rm max}/a_0)\approx 1.5$, where $a_{\rm max}=0.05$~AU,
and $a_0\approx 0.011$~AU is the initial separation of a typical WD binary that merges in 10~Gyr (our assumed age of the Galaxy;
see Paper I). Offsets in $\alpha$ between merger rates separated by a decade are also 1.5. Despite the broad range of allowed
$\alpha$ and \fb\ values, the SDSS data put stronger constraints on the specific WD merger rate. The likelihood-weighted merger
rate is $1.4 \times 10^{-13}$ \ym\ for all systems, and $1.0 \times 10^{-14}$ \ym\ for super-Chandrasekhar
systems. Table~\ref{tab-1} gives 95 \% confidence levels on these rates, and allowed ranges of \fb\ for specific values of
$\alpha$.

\section{DISCUSSION AND CONCLUSIONS}
\label{sec:discussion}

\cite{Li2011} have recently measured the SN Ia rate in galaxies of various Hubble types. For Sbc spirals with the stellar mass of
the Milky Way \citep[$\mathrm{6.4 \pm 0.6 \times10^{10}}$ \msun,][]{McMillan2011} the specific SN Ia rate is $1.1 \times 10^{-13}$
SNe $\mathrm{yr^{-1}\,M_{\odot}^{-1}}$. This number is remarkably close to our measured specific WD merger rate, but an order of
magnitude higher than the super-Chandrasekhar merger rate.

The implication is that there are not enough super-Chandrasekhar DD systems in our local region of the Milky Way to reproduce the
measured SN Ia rate through the classic DD channel. Several authors have already pointed out the apparent dearth of
super-Chandrasekhar SN Ia progenitors using qualitative arguments
\citep[e.g.,][]{Isern1997,maoz08:fraction_intermediate_stars_Ia_progenitors,ruiter09:SNIa_rates_delay_times}. Our analysis shows
this from a quantitative measurement of pre-merger WD binaries. However, we find a remarkable agreement between the total WD
merger rate and the SN Ia rate. For our assumed primary and secondary mass distributions, $\sim$90\% of mergers are
sub-Chandrasekhar, but the total masses are often relatively high -- 10-30\% of the mergers are $>1.2$ \msun, and 25-50\% are
$>1.1$ \msun. Interestingly, recent theoretical work has explored the possibility of sub-Chandrasekhar CO/CO WD mergers leading to
SN Ia \citetext{\citealp{VanKerkwijk2010}, see also \citealp{Guillochon2010,Pakmor2011}}. Apart from the consistency between SN Ia
rates and total WD merger rates, sub-Chandrasekhar explosions may have the advantage of producing the correct chemical
stratification \citep{Sim2010}, without resorting to the ad hoc delayed detonation mechanism \citep{khokhlov91:ddt} needed by
super-Chandrasekhar models. The measured WD merger rate is also important for estimatimg `foregrounds' for gravitational wave
detectors \citep{nelemans09:binaries_review}.

Unavoidably, our conclusions depend to some degree on the assumptions made in the analysis. First, all WD samples are flux
limited, which could lead to selection effects and unrepresentative values of \fb, $\alpha$, or the WD mass distribution. For
example, a DD population with a high super-Chandrasekhar merger rate may exist, but remain unobserved. Even so, the DD population
we observe, with its high merger rate, does exist, regardless of flux limits. Second, we have modeled the initial WD separation
distribution with a power law that is independent of mass. The true post-common-envelope distribution could have some preferred
scale correlated with the component masses, which would affect the parameters and merger rates deduced from the \drvm\
distribution. We have chosen primary masses based on the distribution observed by \cite{kepler07:WD_mass_distribution} for SDSS
WDs, which are mostly single, and we have drawn secondary masses from a flat mass-ratio distribution. These distributions might be
different in DD systems. Different mass distributions would only affect weakly the binary population parameters and the total
merger rate, but they would have a strong effect on the super-Chandrasekhar fraction.  We might have underestimated \drvm\ for
some double-lined WD binaries that are unresolved at the SDSS spectral resolution, and this would lead to an underestimated merger
rate.  Conversely, some of the high \drvm\ systems could conceivably be due to faint M star, rather than WD, companions. However,
an M star of low enough mass to go undetected in the SDSS photometry, $\lesssim 0.15$~M$_\odot$, and with a period of $\gtrsim 1$
hour, would only induce a small \drvm\ in the primary WD, e.g., 220 or 270 km~s$^{-1}$, for WDs of 0.6 M$_\odot$ and 0.4
M$_\odot$, respectively. Such systems therefore cannot dominate the high-\drvm\ tail. At the spectral resolution of SDSS, the
combined absorption plus emission line profiles of some WD+M binaries could mimic a large \drvm, but this should not occur
identically for several Balmer lines, so these systems would be removed by our vetting procedure.  Finally, we have assumed that
the velocity errors are normally distributed, with standard deviations quantified by our empirical tests, but there might be some
low, non-Gaussian, tails to the error distribution that contaminate the \drvm\ distribution with false positives. In upcoming
work, we will address these caveats by investigating the effects of different input assumptions on the conclusions. We will also
improve the purity of our \drvm\ distribution with spectroscopic follow-up of a large number of binary candidates, and use this
improved distribution to refine our constraints on the Galactic DD population and its merger rate.

\begin{deluxetable*}{lccc}
  \tablewidth{10cm}
  \scriptsize
  \tablecaption{Local WD Merger Rates and 95\% Confidence Limits \label{tab-1}}
  \tablecolumns{4}
  \tablehead{
    \colhead{} &
    \colhead{} &
    \colhead{Total rate} &
    \colhead{Super-Ch rate} \\
    \colhead{$\alpha$} &
    \colhead{\fb} &
    \colhead{($10^{-13}$ \mym)} &
    \colhead{($10^{-13}$ \mym)} 
  }
  \startdata
  entire range & $0.014$ to $0.32$ & $1.4$ ($0.16$, $7.2$) & $0.1$ ($0.016$, $0.4$) \\
  $1.0$ & $0.11$ to $0.24$ & $0.3$ ($0.065$, $0.5$) & $0.03$ ($0.017$, $0.045$) \\
  $0.0$ & $0.046$ to $0.22$ & $1.0$ ($0.46$, $2.2$) & $0.08$ ($0.03$, $0.16$) \\
  $-1.0$ & $0.021$ to $0.11$ & $3.0$ ($1.0$, $6.0$) & $0.16$ ($0.05$, $0.3$) %\\
  %$-3.0$ & $0.015$ to $0.034$ & $14$ ($9.0$, $20$) & $1.1$ ($0.74$, $1.8$) 
  \enddata
\end{deluxetable*}

\acknowledgements{We are indebted to Steve Bickerton for his contribution to DS/DT, and to Scot Kleinman for making his
  WD catalog available to us in advance of publication. We acknowledge useful discussions with Tim Beers, Steve
  Bickerton, Joke Claeys, Brian Metzger, Ehud Nakar, Gijs Nelemans, and Marten van Kerkwijk. We are grateful to the
  SWARMS team: Mukremin Kilic, Tom Matheson, Fergal Mullally, Roger Romani, and Susan Thompson. This work was supported
  by a grant from the Israel Science Foundation, and IRG grant number 276988 from the European Union.

  Funding for the SDSS and SDSS-II has been provided by the Alfred P. Sloan Foundation, the Participating Institutions,
  the National Science Foundation, the U.S. Department of Energy, the National Aeronautics and Space Administration, the
  Japanese Monbukagakusho, the Max Planck Society, and the Higher Education Funding Council for England. The SDSS Web
  Site is http://www.sdss.org/.

  The SDSS is managed by the Astrophysical Research Consortium for the Participating Institutions. The Participating
  Institutions are the American Museum of Natural History, Astrophysical Institute Potsdam, University of Basel,
  University of Cambridge, Case Western Reserve University, University of Chicago, Drexel University, Fermilab, the
  Institute for Advanced Study, the Japan Participation Group, Johns Hopkins University, the Joint Institute for Nuclear
  Astrophysics, the Kavli Institute for Particle Astrophysics and Cosmology, the Korean Scientist Group, the Chinese
  Academy of Sciences (LAMOST), Los Alamos National Laboratory, the Max-Planck-Institute for Astronomy (MPIA), the
  Max-Planck-Institute for Astrophysics (MPA), New Mexico State University, Ohio State University, University of
  Pittsburgh, University of Portsmouth, Princeton University, the United States Naval Observatory, and the University of
  Washington.}

%\bibliography{/Users/carles/Documents/library}

\end{document}